\documentclass{iaus}
\usepackage{graphicx}

\title[Mrk~1419 - a new distance determination] 
{Mrk~1419 - a new distance determination }

\author[C.M.V. Impellizzeri {\it et al. } ]  
{C. M. Violette Impellizzeri$^{1,2}$, James A. Braatz$^1$, Cheng-Yu Kuo$^3$,
  Mark J. Reid$^4$,  K.Y. Lo$^1$, Christian Henkel$^5$, James J. Condon$^1$ }
\affiliation{$^1$ National Radio Astronomy Observatory, 520 Edgemont Road,
  Charlottesville, USA \\ email: {\tt vimpelli@alma.cl} \\[\affilskip]
  $^2$ Joint Alma Observatory, Al\'onso de Cordova, Vitacura,  Santiago, Chile  \\[\affilskip]
  $^3$ Institute of Astronomy and Astrophysics, Academia Sinica, Taipei 106, Taiwan \\[\affilskip]
  $^4$  Harvard-Smithsonian Center for Astrophysics, 60 Garden Street, Cambridge, USA \\ [\affilskip]
  $^5$ Max-Planck-Institut f\"ur Radioastronomie, Auf dem H\"ugel  69, 53121 Bonn, Germany}

\pubyear{2012}
\volume{287}  
\pagerange{119--126}
\jname{Cosmic Masers - from OH to H$_{0}$}
\editors{R. Booth, E. Humphreys, W. Vlemmings}
\begin{document}

\maketitle

\begin{abstract}

  Water vapor megamasers from the center of active galaxies provide a powerful tool to trace accretion disks at sub-parsec resolution and, through an entirely geometrical method, measure direct distances to galaxies up to 200\,Mpc. The Megamaser Cosmology Project (MCP) is formed by a team of astronomers with the aim of identifying new maser systems, and mapping their emission at high angular resolution to determine their distance. Two types of observations are necessary to measure a distance: single-dish monitoring to measure the acceleration of gas in the disk, and sensitive VLBI imaging to measure the angular size of the disk, measure the rotation curve, and model radial displacement of the maser feature.  The ultimate goal of the MCP is to make a precise measurement of H$_{0}$ by measuring such distances to at least 10 maser galaxies in the Hubble flow.  We present here the preliminary results from a new maser system, Mrk~1419.  Through a model of the rotation from the systemic masers assuming a narrow ring, and combining these results with the acceleration measurement from the Green Bank Telescope, we determine a distance to Mrk~1419 of 81$\pm$10\,Mpc.  Given that the disk shows a significant warp that may not be entirely traced by our current observations, more sensitive observations and more sophisticated disk modeling will be essential to improve our distance estimation to this galaxy.

\end{abstract}

\firstsection 
\section{Introduction}

The Megamaser Cosmology Project (MCP) aims to determine the Hubble Constant to within 3\%, by accurately measuring the distance to 10 galaxies in the Hubble flow.  The Hubble Constant is an important complement to CMB data for constraining the nature of Dark Energy, the geometry (flatness) of the universe, and the fraction of the critical density contributed by matter. The technique used by the MCP, first pioneed by Herrnstein et al. (1999) to measure the distance to NGC~4258, uses water megamaser emission at 22\,GHz from the center of active galaxies to trace the inner disk geometries at high angular resolution (mas-scale), and thus determine their angular size. The angular size of this inner disk is then compared to the linear size measured through single dish observations, yielding the distance to the galaxy.  The ability to image these objects at such high angular resolution comes through the very high brightness provided by the maser process, the maser disks observed by the MCP are usually extremely compact, extending to $<< $ 1\,pc from the central black hole, and. The spectral signature of such a maser disk is a cluster of systemic H$_2$O features, and two additional H$_2$O clusters, one red- and one blue-shifted with respect to the cluster of systemic features.  Masers disks suitable for the distance technique require a special geometry (the nuclear accretion disk has to be edge-on for significant maser amplification; Lo, 2005), and therefore are extremely rare, so many galaxies must be surveyed to find good candidates, which are then followed up with VLBI
 
The precision obtained by the MCP in determining H$_0$ depends on the quality of the individual measurements, but also on the number of galaxies that can be measured, their distance distribution, and distribution on the sky.  An overall 3\% precision in H$_0$ therefore can be achieved by measuring the distances to 10 galaxies if each distance could be measured to 10\% precision, assuming the individual distance measurements are uncorrelated.  There are currently about 150 galaxies detected in water vapor maser emission, of which about one third show some evidence of disk origin. NGC 4258 is the only galaxy with a 10\% or better distance determination, but it not suitable to constrain H$_0$ directly as it is too close and could be mostly affected by peculiar motion. On the other hand, the MCP is currently studying in detail six H$_2$O maser disks in galaxies which are well into the Hubble flow (e.g Braatz et al. 2010, Kuo et al. 2011). However, because a broad distribution of megamaser sources in the sky is essential for reducing measurement uncertainties, surveys to find more such galaxies remain crucial for the success of the project.

We present here recent results on Mrk~1419, which is one of galaxies studied within the MCP with the aim of determining an accurate geometric distance. The megamaser in Mrk~1419 was discovered in 2002 by Henkel et al. with the 100\,m Effelsberg telescope, and it was the first maser after NGC~4258 to display the characterestic ``disk signature'', but is ten times farther away. Through single dish monitoring with Effelsberg, the authors could already measure the secular acceleration of the systemic components, and therefore concluded that the maser emission in Mrk 1419 must arise from an almost edge-on circumnuclear disk.

\section{Observations and calibration}

We observed Mrk~1419 with the global VLBI between May 2009 and January 2011, for a total of six epochs of 12 hours each.  The global VLBI array comprises the VLBA,  the GBT, and Effelsberg. In two of these epochs, we also added the EVLA, tuned to the frequency of the systemic masers, in order to improve the signal-to-noise level in this part of the spectrum.

\begin{figure}[ht]
\begin{minipage}[b]{0.5\linewidth}
\includegraphics[scale=0.21]{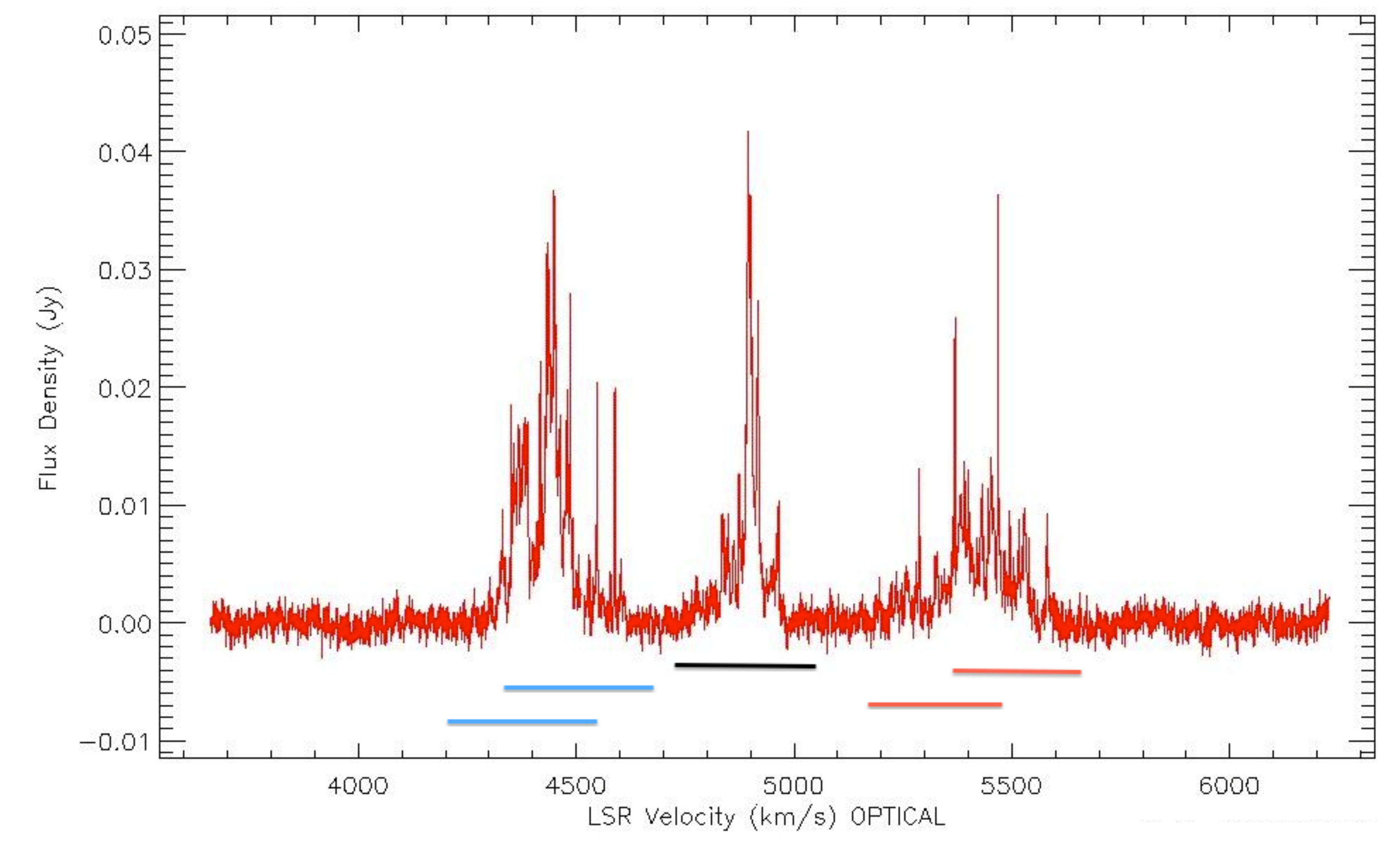}
\end{minipage}
\hspace{0.5cm}
\begin{minipage}[b]{0.5\linewidth}
\includegraphics[scale=0.27]{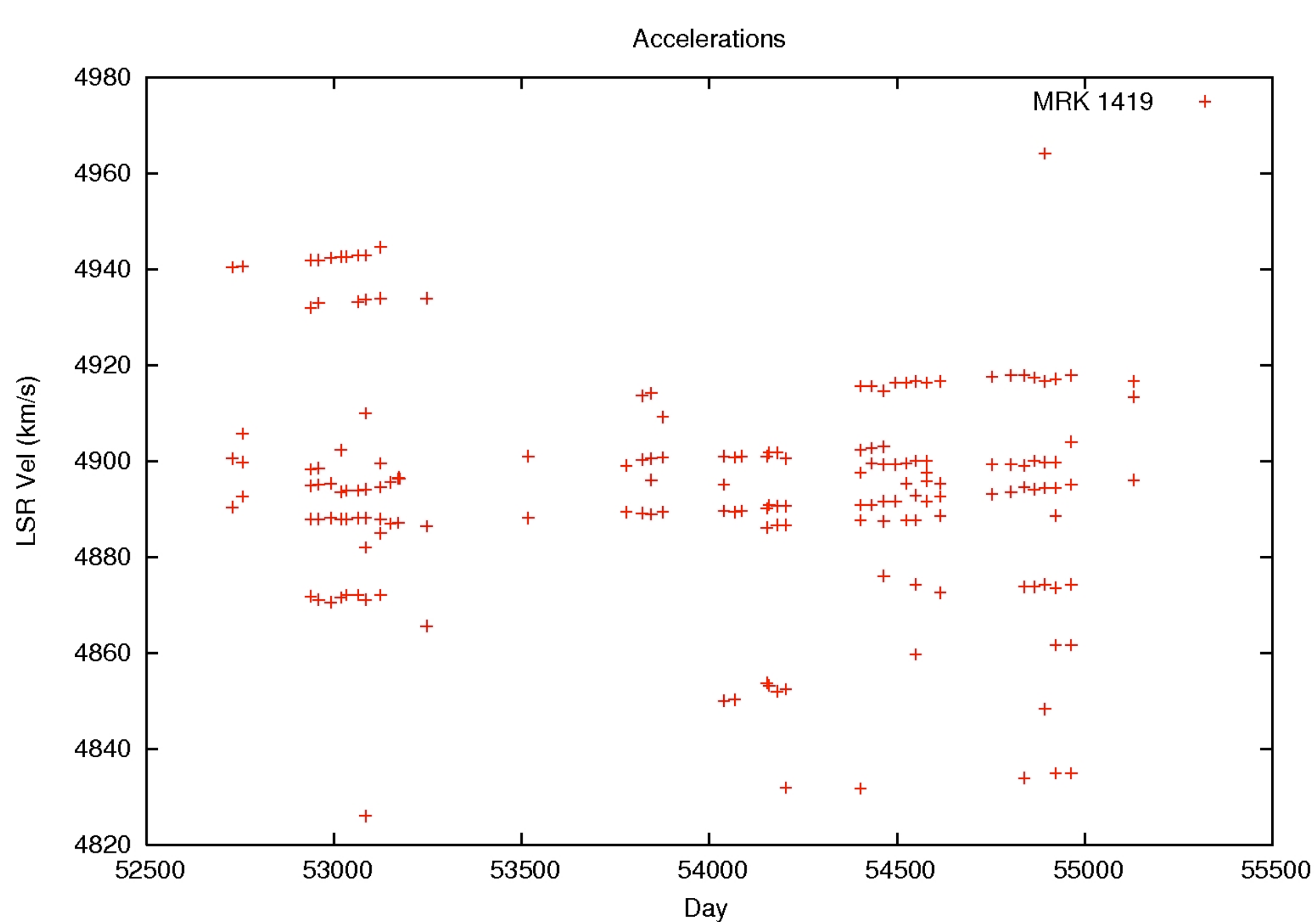}
\end{minipage} \caption{{\it Left}: GBT spectrum of Mrk~1419, taken on December 16, 2009. At the bottem, the position of the VLBI bands are marked for each of the spectral components. Note that for the red and systemic bands, each IF represents two polarizations. {\it Right:} Plot of the systemic maser velocity (on the left shown with the black bar underneath) as a function of time.  The velocity of each maser spot is determined from the GBT observations. The slope in the fit gives the acceleration for each component. } 
 \end{figure}

All our observations were carried out in self-calibration mode.  Figure~1 (left) shows a typical single dish spectrum of Mrk~1419, taken in December, 2009. The systemic masers are typically 40\,mJy, and range over 150\,km\,s$^{-1}$. Given the relative weakness of the masers in this source, and in order to improve the quality of our calibration, we self-calibrated the data using a clump of systemic masers spreading over 10\,km\,s$^{-1}$.  VLBI observations were carried out with four IF bands and two polarizations (RCP and LCP), each of 16\,MHz. Two of the bands were centred on the galaxy systemic velocity, two further were centered on the blue-shifted part of the spectrum, and the last two were centered on the red-shifted part of the spectrum, offset from each other because of the larger spread in velocityin this part (Figure~1, left). "Geodetic" blocks were placed at the start and end of our observations, in order to solve for atmospheric and clock delay residuals for each antenna.

Calibration was performed using AIPS, and included an a priori phase and delay calibration, zenith and atmospheric delays and clock drifts (with the geodetic block data), flux density calibration, a manual phase calibration to remove delay and phase differences among all bands, and selecting a maser feature as the interferometer phase-reference. After calibrating each dataset separately, the data were "glued" together, and imaged in all spectral channels for each of the IF bands. The image from each spectral channel appeared to contain a single maser spot, which we fitted with a Gaussian brightness distribution in order to obtain positions and flux densities.

Single dish GBT monitoring of Mrk~1419 was performed with approximately one observation per month, except for the summer months when the humidity makes observations at 22\,GHz inefficient.  The GBT spectrometer was configured with two 200\,MHz spectral window each with 8192 channels, one centered on the systemic velocity of the galaxy and the second offset by 180\,MHz. Each observation was carried out for about 4\,h.  Finally, data calibration was performed in GBTIDL, with a low order-polynomial fit to the line-free channels to remove the spectral baseline.

\section{Results and discussion }

We present here preliminary results from three VLBI epochs, out of the six epochs observed overall, and from the GBT acceleration measurement performed around those epochs. The VLBI epochs, labeled as BB261N, P and R, were all observed between December 2009 and January 2010. The rms noise level for each of the VLBI maps is $\sim$ 0.8\,mJy\,beam$^{-1}$. Figure~2 (left) shows the maser distribution on the sky, for the three epochs combined, with east-west, and north-south offsets (in mas) relative to the maser components at systemic (black symbols). The position angle of the maser disk is $-131^{\circ}$ and the inclination with respect to the observer is 89$^{\circ}$.  The inner and outer radii of the disk are 0.13 and 0.37\,pc, respectively. The disk shows some warping, especially towards the lower, blue-shifted part. While on the outer part the disk flattens out, there is clear evidence for a significant bending in the inner side.  Towards the red-shifted part of the disk, however, the larger vertical spread in the maser distribution may be due in part to the masers being fainter ($\sim$ 10\,mJy), lowering our signal-to-noise, but may also indicate a true scatter in the maser position, due to a larger inclination in this part of the disk, or may reveal a thicker disk. Figure~2 (right) shows the position-velocity, PV, diagram for Mrk~1419.  The high-velocity masers trace a Keplerian rotation curve and the systemic masers fall on a linear slope. This slope can be extended to the rotation curve traced by the high-velocity features, and this intersection determines the angular radius of the disk and magnitude of the rotation velocity traced by systemic masers. The precise fit to the high velocity masers demonstrate that the disk is dominated by the gravitational potential of the supermassive black hole at the center. From this fit, we calculate the mass of the black hole to be (1.16 $\pm$ 0.05) $\times$ 10$^{7}$\,M$_{\rm solar}$ (for a Hubble constant value of H$_0$ = 73\,km\,s$^{-1}$\,Mpc; see Kuo at al. 2011).

\begin{figure}[ht]
\begin{minipage}[b]{0.5\linewidth}
\includegraphics[scale=0.37]{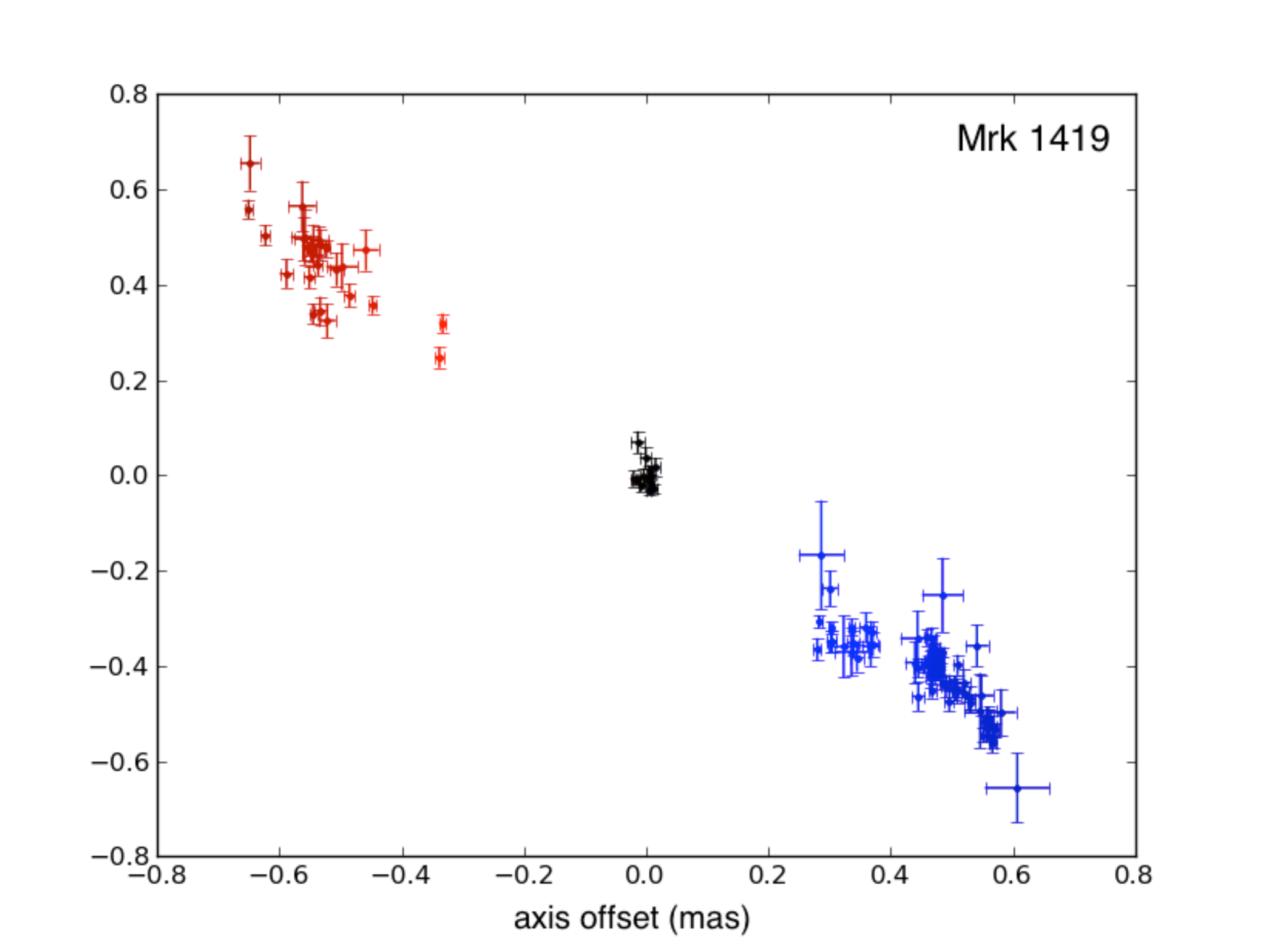}
\end{minipage}
\hspace{0.5cm}
\begin{minipage}[b]{0.5\linewidth}
\includegraphics[scale=0.46]{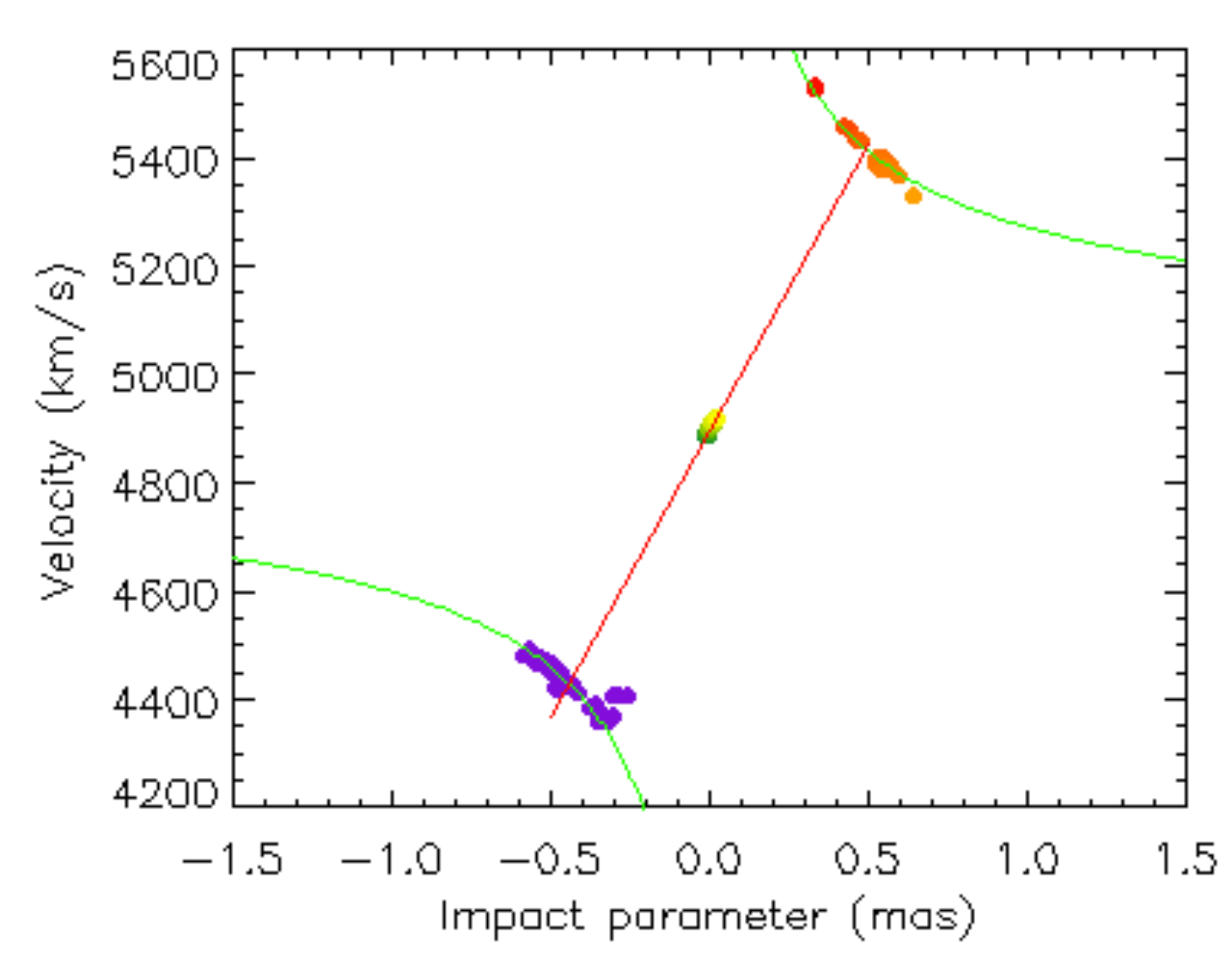}
\end{minipage}
\caption{{\it Left:} Spatial distribution of the inner accretion disk traced by H$_2$O masers. {\it Right:} Position-velocity diagram for Mrk~1419. }
\end{figure}

From the parameters derived from both single dish and VLBI measurements, we measure the angular diameter distance to Mrk~1419 to be 81$\pm$10\,Mpc.  Here, we calculate the distance using: D = a$^{-1}$\,k$^{2/3}$\,$\Omega ^{4/3}$, where a is the acceleration from the GBT results (Figure~1, right), k is the Keplerian rotation constant, derived from the fit to the high velocity masers in the PV-diagram, and $\Omega$ is the slope velocity/impact parameter for the systemic masers (see Braatz et al. 2010). In our most simplified model of the disk, the maser emission originates in a thin, flat, edge-on disk and the dynamics are dominated by a central massive object, with all maser clouds in circular orbits. In this model, high-velocity masers trace gas near the tangential point at the edge of the disk. Systemic masers occupy part of a ring orbiting at a single radius and covering a small range velocities on the near side of the disk. The positive slopes seen in the maser velocities with time (Figure 1, right) indeed show clear evidence for the centripetal acceleration of masers, as they move across the line of sight in front of the central black hole. Using a ``by-eye'' method from similar plots, we measure two accelerations for the systemic masers, one for the masers with velocities $>$ 4940\,km\,s$^{-1}$, of 3.5\,km\,s$^{-1}$\,yr$^{-1}$, and one for the masers $<$ 4940\,km\,s$^{-1}$, of 2.1\,km\,s$^{-1}$\,yr$^{-1}$. Because the higher acceleration masers are fainter and not visible in our VLBI maps, we only take the lower acceleration into account for the distance estimation.  While it is clear that with more than one acceleration the systemic masers likely originate from more than one radial distance from the black hole, as in our simplified assumption, more sensitive VLBI observations will be extremely important in the future to better constrain our models with the available information.  Finally, we determined the slope for the systemic masers in the PV-diagram, $\Omega$, by rotating the disk on the sky by 45$^{\circ}$ (counterclockwise in Figure~2, left) and measured the impact parameter for each maser component by its abscissa on the rotated axes. This method worked well, but has the caveat that the linear slope in the PV diagram of Mrk~1419 is best fitted when the systemic features are rotated by a smaller angle than the disk further out, thus giving further evidence for the presence of a significant warp in the part of the map that is not directly traced by the masers.

\section{Summary and future work}

We presented here VLBI images and single-dish GBT results of the water vapor masers in Mrk~1419. The spatial distribution of the masers in this source is nearly linear, with high-velocity masers on both sides of the masers at the galaxy systemic velocity. The water masers trace gas in Keplerian orbits at radii of $\sim$ 0.2\,pc, moving under the influence of a $\sim$ 1.16 $\times$ 10$^{7}$\,M$_{\rm solar}$ black hole.  We model the rotation from the systemic masers assuming a narrow ring, and combine our results with the acceleration measurement from single dish observations to determine a distance to Mrk~1419 of 81$\pm$10\,Mpc. The main source of uncertainty in the distance comes from the measurement of the orbital curvature parameter $\Omega$, and the uncertainty in the acceleration, while the contribution from the Keplerian rotation constant is negligible. However, the complex geometry in this source is evident from a significant warp in the disk, and the presence of more than one ring for the systemic masers. A more sophistcated modeling of the maser disk using a Bayesian fitting will therefore help solve these complications, and some first results using this method look promising (see Figure~3). Finally, the addition of more sensitive VLBI epochs to our analysis will improve the signal-to-noise ratio and can reduce the distance uncertainty to about 10\%.

\begin{figure}[ht]
\begin{minipage}[b]{0.5\linewidth}
\includegraphics[scale=0.29]{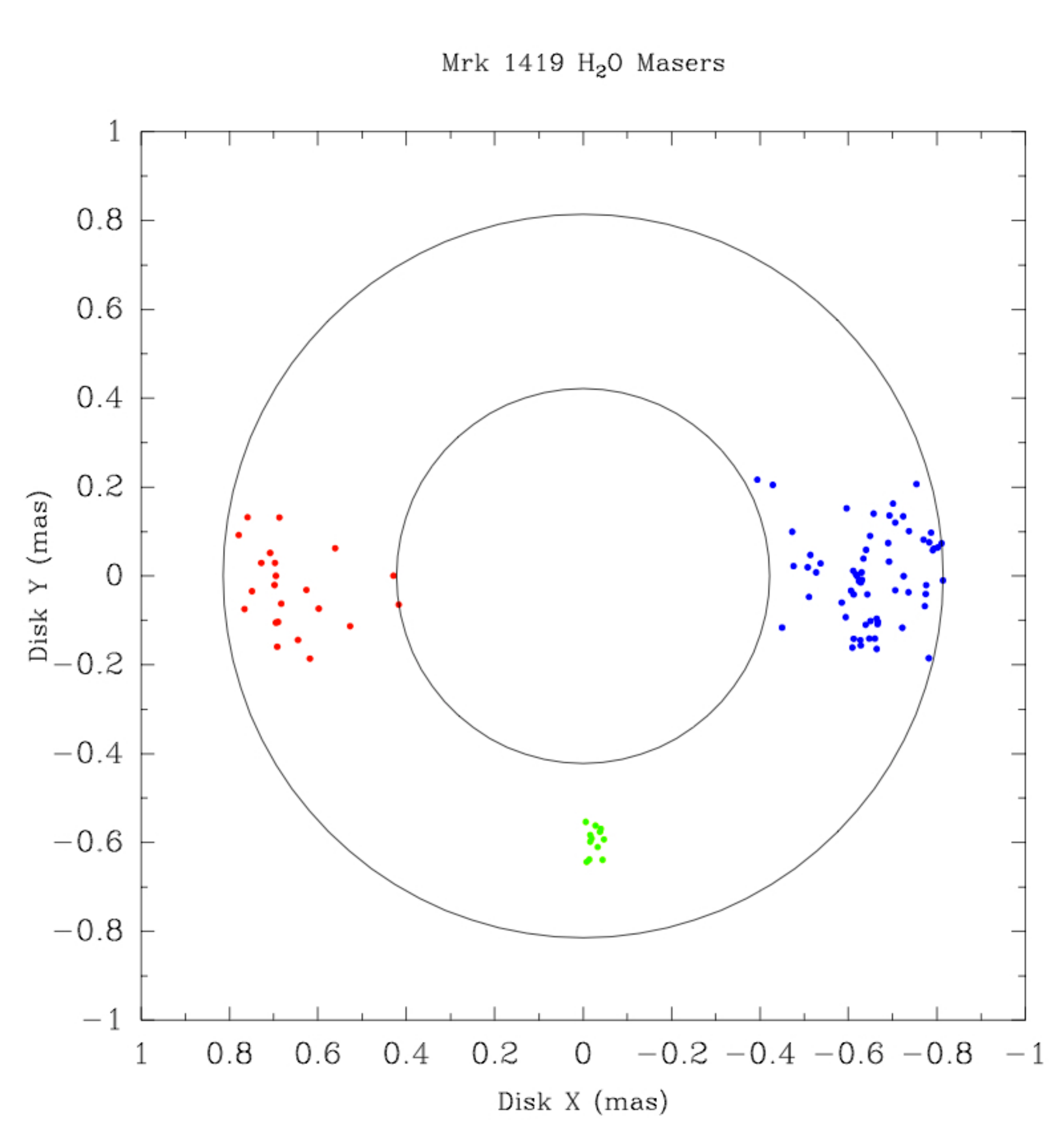}
\end{minipage}
\hspace{0.5cm}
\begin{minipage}[b]{0.5\linewidth}
\includegraphics[scale=0.29]{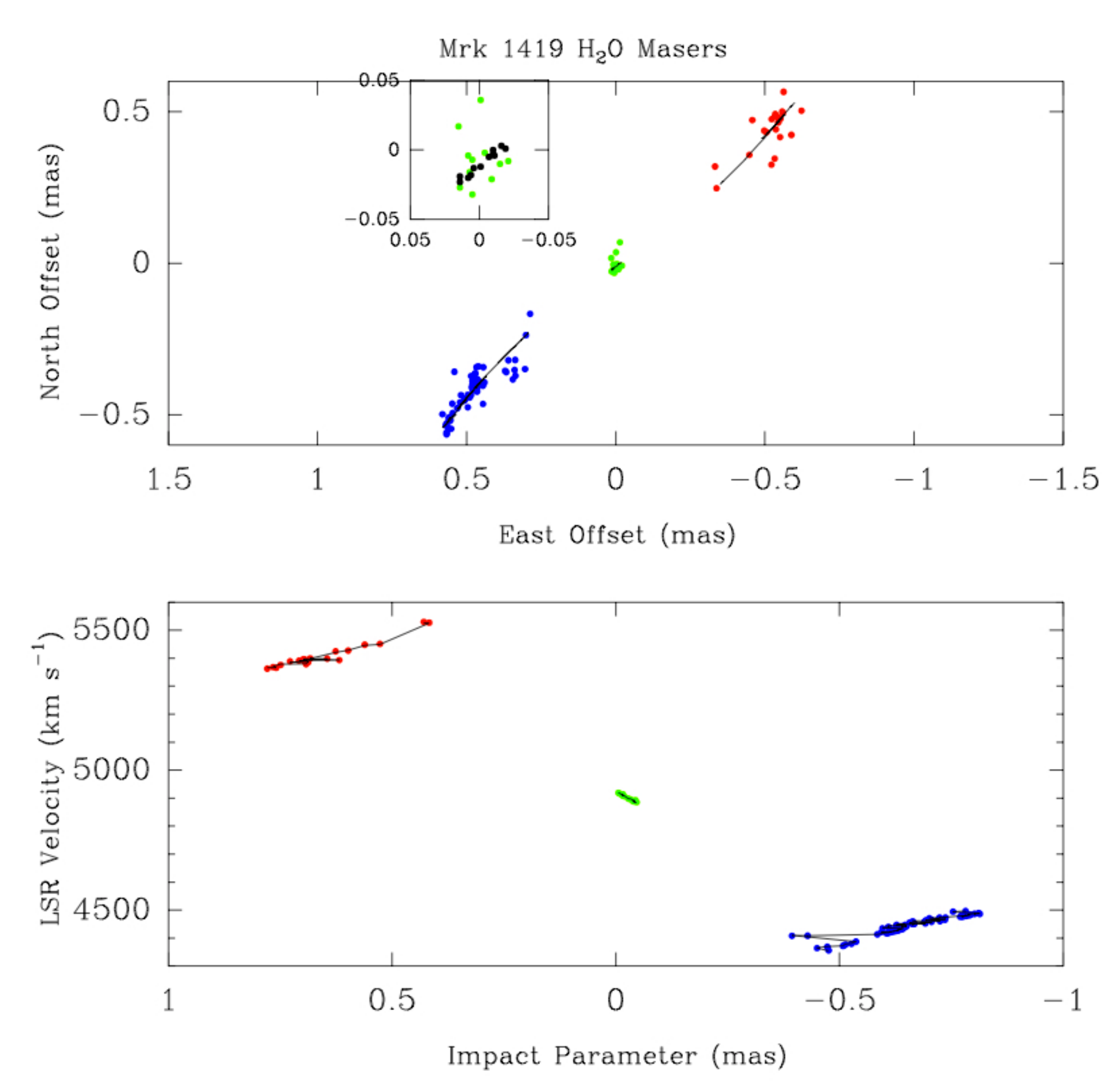}
\end{minipage}
\caption{ {\it Left:} Position of the masers seen from ``above'' the disk, determined from the output of the Bayesian fitting program. {\it Right: } Comparison between model and data. }
\end{figure}

\end{document}